\documentclass[useAMS]{mn2e}
\usepackage{comment}
\usepackage{url}
\usepackage{color}
\usepackage{subfloat}
\usepackage{graphicx}
\usepackage{caption}
\usepackage{amsmath}
\usepackage{breqn}

\def\apj{{\rm ApJ}}

\def\apjs{{\rm ApJS}}

\def\mnras{{\rm MNRAS}}

\def\etal{{\rm et al.~}}

\def\simlt{\lower.5ex\hbox{$\; \buildrel < \over \sim \;$}}
\def\simgt{\lower.5ex\hbox{$\; \buildrel > \over \sim \;$}}

\def\bi{\bibitem[]{}}

\title[Predicting galaxy ellipticities]{
Prediction of galaxy ellipticities and reduction of shape noise in cosmic
shear measurements
}

\author[R.~A.~C. Croft et al.]{Rupert A.~C.~Croft,$^1$\thanks{E-mail: rcroft@cmu.edu} Peter E.~Freeman,$^2$,  Thomas S.~Schuster$^3$ and Chad M.~Schafer$^2$\\
$^1$McWilliams Center for Cosmology, Department of Physics, 
Carnegie Mellon University, Pittsburgh, PA 15213, USA\\
$^2$Department of Statistics, Carnegie Mellon University, 5000 Forbes Avenue, Pittsburgh, PA 15213, USA\\
$^3$Department of Engineering Sciences, University of California Berkeley,
230 Bechtel Engineering Center, Berkeley, CA 94720, USA\\
}

\begin{document}

\pagerange{\pageref{firstpage}--\pageref{lastpage}} \pubyear{2015}

\maketitle

\label{firstpage}

\begin{abstract}
The intrinsic scatter in the ellipticities of galaxies about 
the mean shape, known as ``shape noise,''
is the most important source of noise in 
weak lensing shear measurements. Several 
approaches to reducing shape noise have 
recently been put forward, using information beyond
photometry, such as radio polarization and optical spectroscopy. 
Here we investigate how well the intrinsic ellipticities
of galaxies can be predicted using other, exclusively
photometric parameters. These parameters (such as 
galaxy colours) are already available in 
the data and do not necessitate additional, often expensive observations.
We apply  two regression techniques, generalized additive models (GAM)
and projection pursuit regression (PPR) to the publicly released data 
catalog of galaxy properties from CFHTLenS. 
In our simple analysis we find that the individual galaxy
ellipticities can indeed be predicted
from other photometric parameters to better precision than the
scatter about the mean ellipticity. This means that without
additional observations beyond photometry
the ellipticity contribution to the shear
can be measured to higher precision, comparable to using a larger sample of 
galaxies. Our best-fit model, achieved using PPR,
yields a gain equivalent to having 114.3\% more galaxies.
Using
only parameters unaffected by lensing (e.g.~surface brightness, colour),
the gain is only $\approx 12\%$. 
\end{abstract}


\section{Introduction}
\label{intro}

Weak gravitational lensing of galaxies is the distortion of 
galaxy shapes and sizes viewed behind a distribution of
gravitating matter (see e.g.~Bartelmann \& Schneider 2001 or Massey,
Kitching \& Richard 2010 for reviews). 
The change in 
galaxy shapes, known as cosmic shear, has
become one of the main probes of cosmology due to its dependence on 
the total matter distribution and cosmic geometry (e.g.~Kaiser 1998).
It is a driver for many ambitious upcoming instruments, including
LSST,\footnote{http://www.lsst.org} Euclid,\footnote{http://sci.esa.int/euclid}
 and WFIRST (Spergel et al.~2015). On an individual galaxy
basis, cosmic shear induces changes in the ellipticity and position
angle at the few percent level. Determination of cosmic
shear therefore relies on measuring coherent distortions, averaging over large
numbers of galaxies. The dominant source of noise in this measurement
is so-called ``shape noise,'' which is due to the fact that the unlensed
galaxies have an intrinsic distribution of ellipticities and
position angles. This distribution must be averaged over to reveal the 
cosmic shear contribution. If the shapes and position angles of
the unlensed galaxies were known, then this shape noise could be 
eliminated, and consequently many fewer galaxies would be needed to achieve 
a given precision in cosmic shear.

This realization has given rise to several proposed techniques to determine
the unlensed shapes of individual galaxies, using additional observables
beyond photometry. The most prominent idea is to use spectroscopic
information to do this.
Early work on this subject involved spatially resolved kinematic maps
of galaxies (Blain 2002, Morales 2006).
 Recently, Huff \etal (2013) have shown than the 
disk galaxy line width-luminosity relationship 
(Tully \& Fisher 1977) can in principle
be used to elminate shape-noise as an important source
of noise altogether. This is extremely
promising, and these authors have shown that spectroscopic lensing survey
concepts can be conceived which are significantly smaller in scale
than LSST but which are highly competitive in terms of 
predicted dark energy constraints. Other recent work by 
Brown \& Battye (2011) has shown how polarization angles measured from
radio observations can yield intrinsic galaxy positions angles.
Again these techniques require the use of additional information
beyond galaxy photometry.

There is information in photometry itself however on the intrinsic
shapes of galaxies. For example, there are well-known relationships
between the inclination angles of galaxy disks and their surface
brightnesses (e.g. Giovanelli et al.~1994).
 One could imagine measuring the surface brightness
(which is unaffected by lensing)
from images and then using this relationship to infer something
about the unlensed shape. In this paper, we will extend this idea
to all photometrically measurable information, and apply it
to a published observational dataset, the 
Canada-France-Hawaii Telescope Lensing Survey (CFHTLenS; Heymans et al.~2012).
The question we will try to answer is: is it possible to reduce the
shape noise in weak lensing shear without resorting to extra observables
beyond photometry (which are often expensive to obtain)?
 
There are many photometric variables which can be measured from 
galaxy images (which often includes colour information from 
different filters). Some of these variables
are affected by lensing (for example the
size or the apparent magnitude of galaxies) and others not (such as the
surface brightness or the photometric redshift). There will be correlations
between these variables and the intrinsic ellipticity of galaxies,
and in this paper we will investigate how these correlations can be
used to predict the intrinsic ellipticity. 
We will be using a set of 
16 parameters for each galaxy taken from those
measured and published by the CFHTLenS team
(see Section \ref{cfhtlens}
 for details). Because there are a large number of parameters,
we will not look at correlations of each individually with galaxy ellipticity,
but instead use two regression techniques, generalized additive models (GAM)
and projection pursuit regression (PPR),
to optmize ellipticity prediction in the multidimensional
parameter space. 

The outline of this paper is as follows: in Section 2 we briefly outline
how galaxy ellipticities are defined and can be used to infer the shear
due to weak lensing. In Section 3 we introduce the data from CFHTLenS, and 
in Section 4 we describe our method for predicting ellipticities from 
other photometric parameters. In Section 5 we present our results for
how well the ellipticities can be predicted, using observational data. 
We summarise and discuss our findings in Section 6.

\section{Weak lensing shear and ellipticity measurements}

We note that the intrinsic ellipticities 
of galaxies are of course not available in the CFHTLenS dataset,
and so when making predictions for them, we will 
compare the predictions to the actual measured ellipticities. Because the
effects of weak lensing shear on the ellipticities are galaxies
are much smaller (around the percent level) than the error on the
predicted ellipticities, this will be a good approximation
to comparison to the intrinsic ellipticities.

The weak gravitational lensing shear can be decomposed into
the two usual components, sometimes denoted
as $\gamma_{\times}$ and $\gamma_{+}$, which distort the position angle
and ellipticity of the galaxy image. The ellipticity $e$, of a galaxy image
is given by 
\begin{equation}
e=\left(\frac{1-q^{2}}{1+q^{2}}\right)
\end{equation}
where $q=b/a$, the ratio of minor to major axis.
The distortion caused by the shear means that the observed value of $q$ for
a galaxy is given by 
\begin{equation}
q_{\rm obs}=q_{\rm unlensed}(1+2\gamma_{+}),
\end{equation}
with the position angle, $\theta$, changing as follows:
\begin{equation}
\theta_{\rm obs}=\theta_{\rm unlensed}+\frac{\gamma_{\times}}{e},
\end{equation}
where $q_{\rm unlensed}$ and $\theta_{\rm unlensed}$ are galaxy
parameters before lensing distortion. 
It follows that an estimator for the shear component
$\gamma_{+}$ could be
\begin{equation}
\hat{\gamma}_{+}=\frac{1}{2}\left(\frac{q_{\rm obs}}{q_{\rm unlensed}}-1\right).
\end{equation}
which would necessitate knowledge of the unlensed galaxy shape,
$q_{\rm unlensed}$. In this paper, we will see whether the unlensed
ellipticity can be inferred from other parameters. We leave the determination
of a shear estimator that makes best use of this information to future
work.

We note that in our work we will not be able to
infer the unlensed position angles. There will therefore be no
reduction of shape noise for one of the two shear components, 
$\gamma_{\times}$.
This is likely to be a significant limitation, as for example
Whittaker et al.~(2014) have shown that shear estimators can be constructed
using galaxy position angles only, and which appear to contain most of
 the shear
signal.

\section{Data}

\label{cfhtlens}

We use the publicly available 
data\footnote{http://www.cfhtlens.org/astronomers/data-store}
from CFHTLenS in our analysis.
CFHLenS is a 154 square-degree multi-colour optical survey in {\it ugriz}
 incorporating all five years worth of data from the Wide, 
Deep and Pre-survey components of the CFHT Legacy 
Survey.\footnote{http://www.cfht.hawaii.edu/Science/CFHTLS/}
The CFHTLenS was optimised for weak lensing analysis with the deep
$i$-band data taken in optimal sub-arcsecond seeing conditions.
For a general overview of the survey see Erben et al.~(2013) and Heymans 
et al.~(2012), as well as information about the photometry in
Hildebrant et al.~(2012).

The online datastore$^{3}$ contains 107 photometrically
derived parameters for each of 8.05 million galaxies,
ranging from the number of exposures, through galaxy
 angular positions and photometric redshifts, image
ellipticity components and point spread functions.
The algorithm {\tt lensfit} (Miller et al.~2007) was used by
Miller et al.~(2013) to carry out the Baysian estimation of the galaxy
shapes. 
%The data were downloaded from the CFHTLenS website 
%\footnote{\texttt{http://www.cfhtlens.org/}}.  
%To separate stars from galaxies, we used the parameter \texttt{CLASS_STAR}, 
%drawn from the SExtractor processing  \citep{1996A&AS..117..393B}.  
%This is a neural network-based classifier based on isophotal areas and peak 
%intensity.  It assigns each object an index between 0 (certain galaxy) 
%and 1 (certain star); we selected galaxies as objects with 
%\texttt{CLASS_STAR}$<0.5$.
The {\tt lensfit} shapes require a multiplicative (Miller et al.~2012) and an 
additive (Heymans et al.~2012) correction to be properly calibrated.
Corrections were made (specifically to the measurement {\tt e2}) prior
to the start of our analysis (M.~Simet, private communication).
%One can make the additive correction on an object-by-object basis
From the possible pool of predictor variables, we concentrate on a 
particular subset of 16, which we list in Table \ref{ptable}.
The reader is referred to the CFHTLens publications 
and catalogue documentation (both listed above) for detailed
definitions of these parameters and explanations of how they
were measured. Note that we do not a priori expect two of these parameters
({\tt t\_ml} and {\tt t\_b}) to have any predictive power in estimating
galaxy ellipticity; in a sense, these are control variables added to ensure
proper performance of our analysis software/algorithm.

%We have chosen a subset of 25 parameters to include in our analysis, 
%and will test the
%effect of dividing these parameters further. The parameters we
%have chosen to use are listed in Table \ref{ztable}.

%We compare the galaxy ellipticties we predict from
%other photometric parameters to the {\tt lensfit} measurements of the
%observed galaxy ellipticities, also from CFHTLenS.
%We compute the mean
%value of the measure galaxy ellipticities 
%in CFHTLenS and find it to be $\langle e \rangle = 0.3356$.
%The $rms$ deviation from the mean,
%\begin{equation}
%e_{\rm rms}= \sqrt{{\textstyle \sum}^{N_{\rm gal}}_{i=1} 
%(e_{i} -\langle e \rangle)^{2}/N_{\rm gal} } =0.1722,
%\end{equation}
%where $e_{i}$ is the ellipticity of galaxy $i$
%and the sum is over all $N_{\rm gal}=8.05$ million galaxies in the sample.

For computational efficiency, we select data spanning 100{,}000 
contiguous rows of the catalog, then exclude those whose measurements
include the values {\tt 99} or {\tt -99}, those classified as stars with
high probability ({\tt class\_star} $>$ 0.95), and those for which
{\tt (fitclass | star\_flag)} $\ne$ 0. The final sample size is 89{,}990,
which is sufficiently large to probe the space spanned by the predictor
variables. We compare galaxy ellipticties that we predict from
other photometric parameters to the observed values from {\tt lensfit}
(i.e.~{\tt ellip} in Table \ref{ptable}).
The mean value of {\tt ellip} in our 89{,}990-galaxy sample is
$\bar{e} = 0.3412$, while the root-mean-square (RMS) deviation from the mean is
\begin{equation}
e_{\rm RMS}= \sqrt{\frac{1}{N_{\rm gal}} \sum^{N_{\rm gal}}_{i=1} (e_i - \bar{e})^2} = 0.1759 \,.
\end{equation}
%where $e_{i}$ is the ellipticity of galaxy $i$
%and the sum is over all $N_{\rm gal}=8.05$ million galaxies in the sample.

\section{Analysis}

Our interest lies in predicting ellipticities as a function of the predictor
variables in Table \ref{ptable}. There are 
many fitting techniques available from the
realms of statistics and machine learning that may be applied to this problem;
we find that a combination of two regression techniques, generalized additive
models (GAM; see e.g.~Chapter 7 of James et al.~2013) and projection pursuit 
projection (PPR; Friedman \& Stuetzle 1981), yields encouraging 
results. In short, we use GAM to select a set of predictors from the pool of 
possibilities in Table \ref{ptable}, without testing for interactions 
(which adds
undue computational complexity within the GAM framework), and then apply PPR, 
which works with linear combinations of predictors, to generalize the GAM 
model.

The GAM model is
\begin{equation}
\hat{e} = \beta_0 + \sum_{j=1}^p \beta_j f_j(x_j) \,,
\end{equation}
where $p$ is the number of predictors and $f_j(x_j)$ is a nonparametrically
smoothed version of the $j^{\rm th}$ predictor $x_j$. (In our analysis, we
use the {\tt gam.fit} function of the {\tt R} package {\tt mgcv}, and
apply smoothing splines to each predictor individually.) 

To avoid overfitting,
we apply a forward-stepwise search, wherein we test add each predictor from the
pool individually to the baseline model, and see which achieves the greatest 
reduction in mean squared error (MSE):
\begin{equation}
MSE = \sqrt{\frac{\sum^{N_{\rm gal}}_{i=1} w_i (\hat{e}_i - e_i)^{2}}{\sum^{N_{\rm gal}}_{i=1} w_i}} \,,
\label{msedef}
\end{equation}
where $e_i$ and $\hat{e}_i$ are the measured and predicted
ellipticities respectively of galaxy $i$ and $w_i$ is that galaxy's weight as
estimated by {\tt lensfit} (Miller et al. 2012).

\begin{figure}
  \begin{center}
    \includegraphics[scale=0.5]{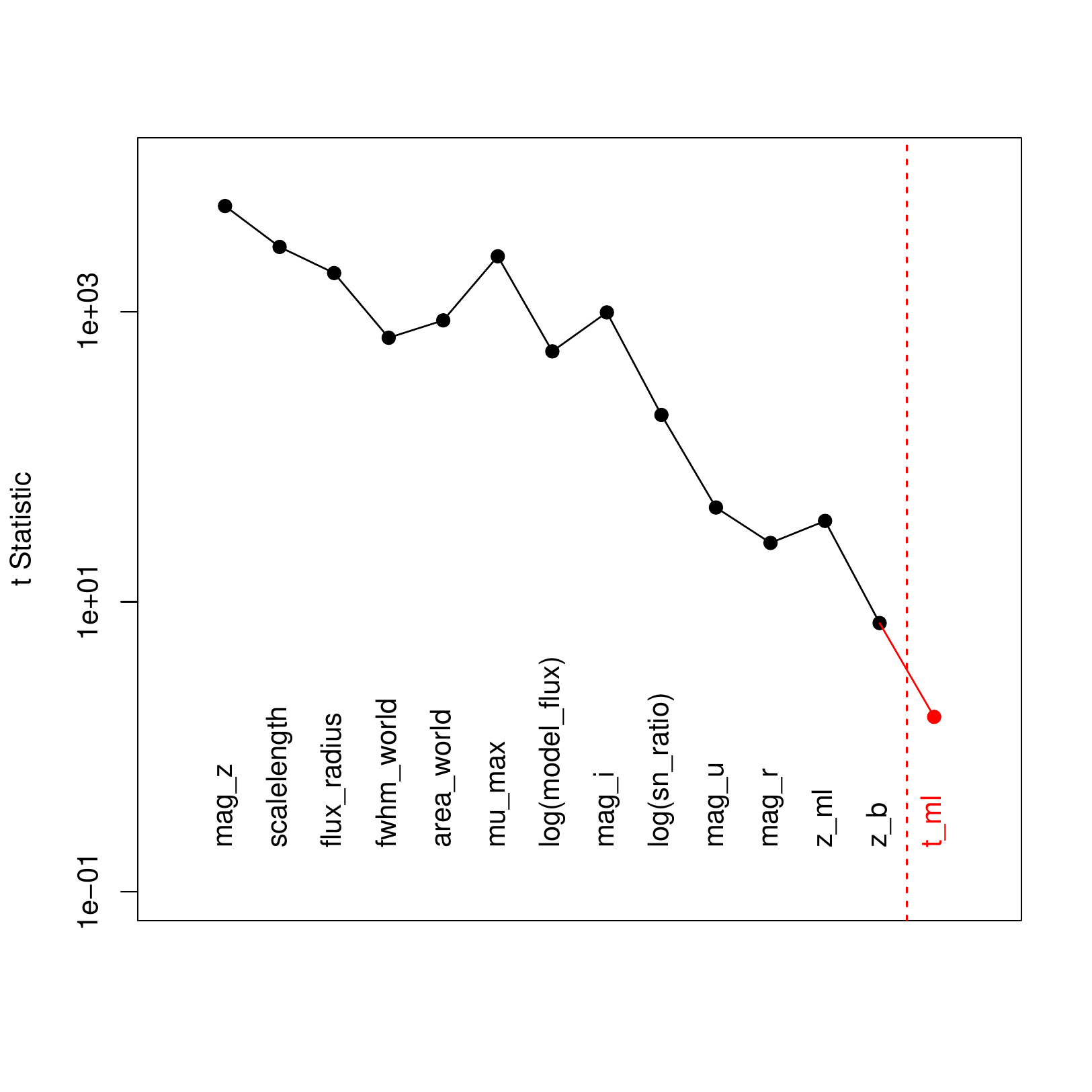}
  \end{center}
  \caption{Result of forward-stepwise model search using generalized 
           additive model regression. Values of the $t$ statistic for the 
           two-sample $t$ test are shown along the y-axis. (See the text for 
           details on how we apply the two-sample $t$ test.) Predictors are
           admitted into the GAM model one at a time, in the order shown from
           left to right. The $p$ value for admitting {\tt t\_ml} is 0.067 and
           thus it was not admitted to the final GAM model.}
  \label{fig:t}
\end{figure}

 To generate predictions $\hat{e}_i$,
we apply five-fold cross-validation, i.e.~we randomly partition the data into
five groups, and at any one time fit our GAM model to four of them to generate
predictions for the fifth group (repeating the process until predictions
are generated for all data). To determine whether the reduction in MSE is
statistically significant, we repeat each fit of each predictor variable
ten times (i.e.~we randomly partition the data into five groups ten separate
times) to generate a distribution of MSEs. Given the MSE distributions for the
baseline and baseline-plus-new-predictor models (which we assume are normal),
it is trivial to apply the two-sample $t$ test to assess the
null hypothesis that the distributions have the same mean. If the $t$ test 
results in a $p$ value $<$ 0.05, we reject the null and incorporate the
new predictor into our baseline model. We then repeat the search over the 
remaining
predictors. Note that as part of this process we check to see if logarithmic
or exponential transformations of the predictors lead to heightened reductions
in MSE. 
We show our results in Figure \ref{fig:t}; the final GAM model, which
includes 13 predictors, reduces the MSE from 0.03424 (the value for a constant
model) to 0.01791. 

Given the set of predictor variables produced in the GAM step, we test 
for interactions among them via PPR. The PPR model is
\begin{equation}
\hat{e} = \beta_0 + \sum_{k=1}^M \beta_k f_k(\boldsymbol{\alpha}_k^T {\bf x}) \,,
\end{equation}
where $f_k(\boldsymbol{\alpha}_k^T {\bf x})$ is the $k^{\rm th}$ 
``ridge function" and
where the number of ridge functions $M$ is selected via the same MSE-reduction
test outlined above. In our analysis, we apply the base {\tt R} function 
{\tt ppr}, and we choose the ``supersmoother" function of Friedman (1984) as
the smoothing function $f_k(\cdot)$.
The final MSE is 0.01622 for $M = 8$; this reduction in ellipticity error
relative to a constant model (MSE = 0.03424) is equivalent to that achieved
using the constant model and a dataset
\begin{equation}
\frac{n_{con}}{n_{ppr}} = \frac{{\rm MSE}_{con}^2}{{\rm MSE}_{ppr}^2} = \frac{0.0324}{0.01622} = 2.111
\end{equation}
times larger, i.e.~111.1\% larger.

\begin{figure}
  \begin{center}
    \includegraphics[scale=0.5]{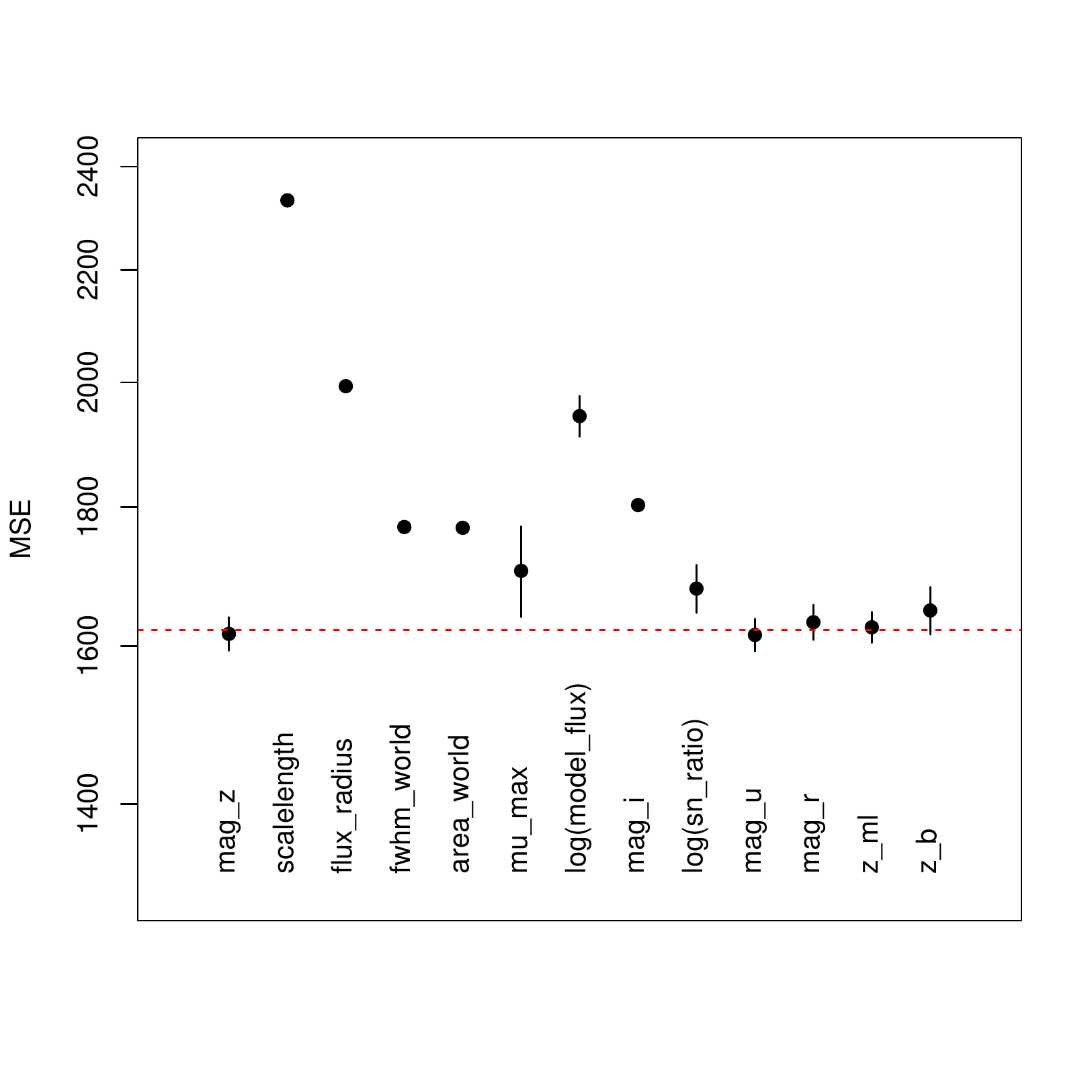}
  \end{center}
  \caption{Mean-squared error (MSE) resulting from the removal of each of the 
           named predictors in turn from the pool of predictors available to
           the PPR model. The red dashed line indicates the MSE for the 
           best-fit PPR model, and the error bars are 1$\sigma$ estimates 
           based on 10 repetitions. This figure indicates that by including
           linear combinations of predictors, several predictors that were 
           significant in the no-interaction GAM model 
       ({\tt mag\_z}{,}{\tt mag\_u}{,}{\tt mag\_r}{,}{\tt z\_ml}{,}{\tt z\_b})
           can be excluded in the PPR model.}
  \label{fig:remove}
\end{figure}

In Figure \ref{fig:remove} we examine the effect of removing each of the
predictors in turn on the MSE of the best-fit PPR model. (Note that we do not
attempt to optimize the number of terms in each case but rather assume
$M = 8$.) The predictors are shown left-to-right in the order in which they
were admitted to the no-interaction GAM model. The effect on MSE of each
parameter largely mimics the $t$ statistics associated with those parameters in
Figure \ref{fig:t}, with the suprising exception that {\tt mag\_z}, the first
predictor chosen in the GAM step, can be excluded from the predictor pool
made available in the PPR step with no loss in predictive accuracy.
Similarly, the last four predictors adopted in the GAM step ({\tt mag\_u},
{\tt mag\_r}, {\tt z\_ml}, and {\tt z\_b}), can similarly be dropped from the
pool. This serves to highlight the complexity of statistical model selection:
the GAM step, designed to reduce the number of possible predictors from
$\sim$ 100 available in general (or from the 16 that we pre-selected for
this particular exercise) to a more manageable pool, still ends up selecting
more than is ultimately necessary because it does not take predictor 
interactions into account.

\begin{figure}
  \begin{center}
    \includegraphics[scale=0.5]{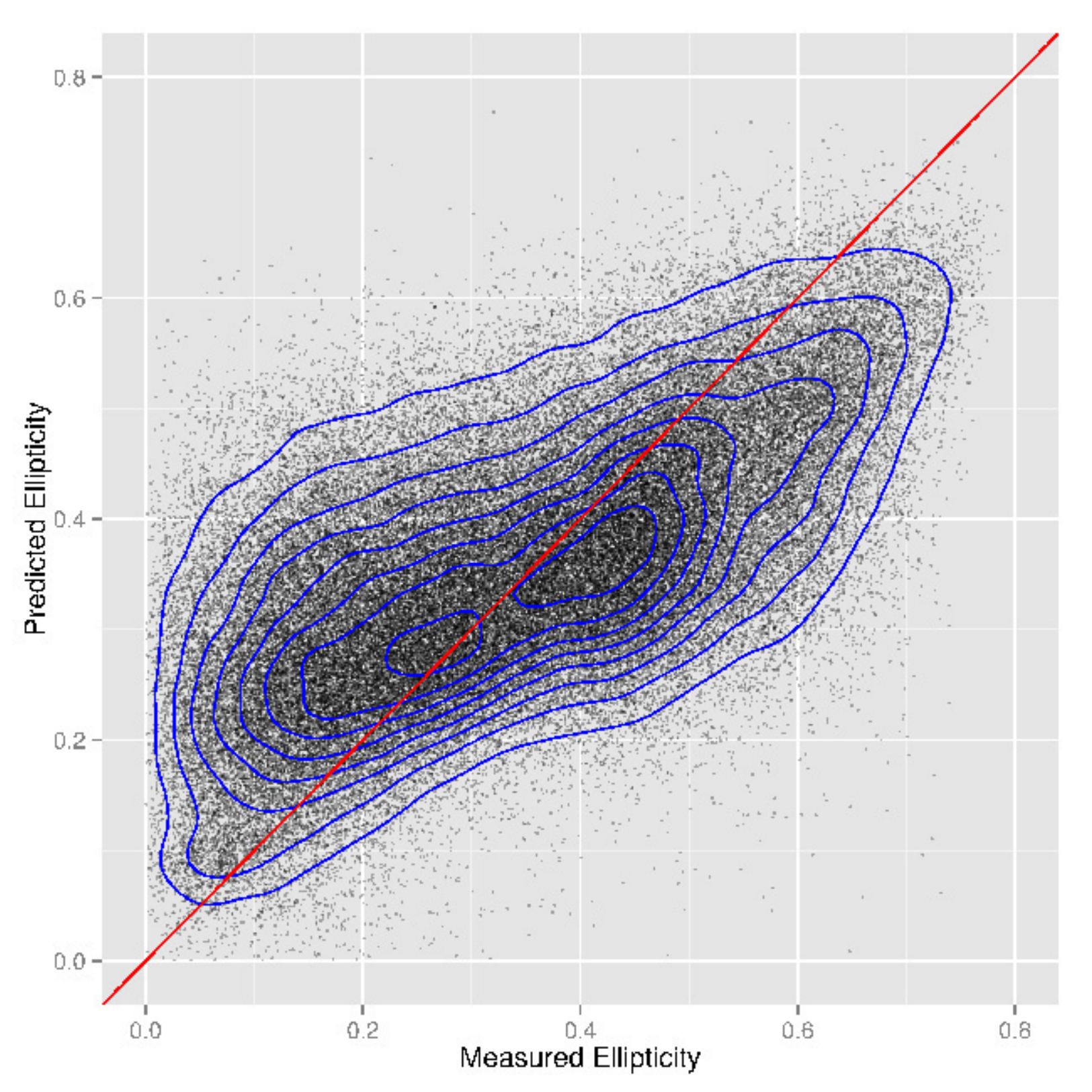}
  \end{center}
  \caption{Measured ellipticity versus predicted ellipticity for the best-fit
           PPR model with MSE = 0.01598 (i.e. $\sigma_e =$ 0.1264). As detailed
           in the text, use of the best-fit PPR model is equivalent to 
           the application of a constant model to a dataset with $\approx$ 110\%
           more galaxies than the current one.}
  \label{fig:fit}
\end{figure}

We thus perform the one additional step of removing the five predictors listed
above from the pool of predictors made available to the PPR model and 
re-running the PPR analysis. We achieve an MSE of 0.01598 for $M$ = 8, which
is equivalent to applying the constant model to a dataset that is 2.143 
times, or 114.3\%, larger. In Figure \ref{fig:fit} we show the relationship 
between predicted and measured ellipticity; the Pearson sample correlation
coefficient between the two is $R = 0.667$.

\begin{table*}
\caption{Variables examined in our GAM-PPR framework. See e.g.~Erben et
al.~(2013), Table C1.}
\label{ptable}
\centering
\begin{tabular}{| l | l |}
\hline
Variable & Description \\
\hline
Predictor Variables:\\
{\tt area\_world} & Galaxy area in world coordinates (= $\pi \times$ {\tt a\_world} $\times$ {\tt b\_world}; \\
 & the latter two quantities are estimated via {\tt SExtractor}) \\
{\tt flux\_radius} & Galaxy half-light radius, estimated via {\tt SExtractor} \\
{\tt fwhm\_world} & Galaxy FWHM assuming a Gaussian profile, estimated via {\tt SExtractor} \\
{\tt mag\_[u,g,r,i,z]} & Galaxy $ugriz$ magnitudes, estimated via {\tt SExtractor} \\
{\tt model\_flux} & Galaxy flux, estimated via {\tt lensfit} \\
{\tt mu\_max} & Galaxy peak surface brightness, estimated via {\tt SExtractor} \\
{\tt scalelength} & Galaxy scalelength, estimated via {\tt lensfit} \\
{\tt snratio} & Galaxy signal-to-noise ratio, estimated via {\tt lensfit} \\
{\tt t\_[b,ml]} & Spectral type, estimated via {\tt BPZ} \\
{\tt z\_[b,ml]} & Galaxy peak-posterior/maximum likelihood photometric redshifts, estimated via {\tt BPZ}\\
\\
Response Variable:\\
{\tt ellip} & Galaxy ellipticity, estimated via {\tt lensfit} (= $\sqrt{{\tt e1}^2 + {\tt e2}^2}$) \\
 & ({\tt e2} corrected by M.~Simet, private communication) \\
\\
Fit Weight:\\
{\tt weight} & Galaxy weight in fitting, estimated via {\tt lensfit}; \\
 & see Section 3.6 and Equation 8 of Miller et al.~(2012) \\
\hline
\end{tabular}
\end{table*}

To illustrate the difference between choosing from all predictors versus only 
those {\em not} affected by lensing, we apply the PPR framework to only
the set of parameters {\tt area\_world}, {\tt mu\_max}, {\tt mag\_(u,r,i,z)},
and {\tt z\_(b,ml)}. We test various combinations of these parameters. First,
we test models with {\tt area\_world} and {\tt mag\_r} and models that
keep information on surface brightness only by combining the two as
{\tt mag\_r}/{\tt area\_world}. Second, we test models incorporating colours
as opposed to magnitudes. Regardless of model, the result is qualitatively
similar: the reductions in MSE relative to that of the constant model are
equivalent to using datasets that are 11.2\%-12.5\% larger, a far smaller
improvement than the 114.3\% gained from examining all predictors.

%In Table \ref{ztable}, we also show the results for predictions based on 
%different parameter sets. The first subset (Nolens parameters, row 2)
%is the set of 13 parameters which are not affected by gravitational
%lensing. These are related to galaxy colour, surface brightness, photometric
%redshift, specifically involving the following parameters in the table:
%(1) ISOAREA\_WORLD (2) MU\_MAX (11) ZB (12) MAG\_u 
%(13) MAG\_g (14) MAG\_r (15) MAG\_i (16) MAG\_y (17) MAG\_z 
%(21) PSF Strehl ratio (22) odds (23) z$_{\rm ML}$ (24) T$_{\rm ML}$.
%In order to keep information on surface brightness only,
%we use MAG r/ISOAREA\_WORLD
%as a parameter instead of the two parameters individually. 
%We also use colour
%information,
%MAG\_g-MAG\_r, MAG\_i-MAG\_r, MAG\_y-MAG\_r, and MAG\_z-MAG\_r, instead
%of the apparent magnitudes.

\section{Summary and discussion}

\subsection{Summary}

%We have used the machine learning technique of K Nearest Neighbours
We utilize a statistical framework based on generalized additive model
(GAM) regression and projection pursuit regression (PPR)
to predict galaxy ellipticities from other photometric parameters, and
apply it 89{,}990 galaxies taken from a value-added version of the public
CFHTLenS catalog.
%We have used the data from CFHTLenS as a training set, and to
%test our method. 
Our findings are as follows:
\begin{enumerate}

\item Using a set of 13 parameters which include quantities which are
affected by lensing such as galaxy size and apparent magnitude, we
find that the ellipticity of individual galaxies can be predicted with
an rms error $\sigma_{e}=0.1264$. This is $28.1\%$ less than the rms standard
deviation of galaxy ellipticities about the mean. The gain in predictive
accuracy relative to a constant model is equivalent to utilizing a constant
model with a dataset 114.3\% larger than our 89{,}990-galaxy CFHTLenS-based
dataset. This result conclusively demonstrates that
our statistical framework can reduce shape noise in weak lensing 
measurements.

\item Using a reduced set of photometric parameters, those unaffected 
by lensing (such as colour and surface brightness), we find that the 
ellipticity of galaxies can be predicted with an rms error of 
$\sigma_{e} \approx 0.1749$, $0.5\%$ less than the rms standard
deviation of galaxy ellipticities about the mean; the gain in predictive
accuracy relative to a constant model is equivalent to utilizing a constant
model with a dataset $\approx$ 12\% larger.

%\item From a jackknife test, we conclude that parameters related to
%galaxy size and luminosity have greater influence in the ellipticity
%prediction than those based on colour. This is consistent with our two 
%findings above.

\end{enumerate}

\subsection{Discussion}
Although we have shown that photometric information can be used
to predict galaxy ellipticities, the scatter compared to the true values
is still large, so that on a galaxy by galaxy basis, photometric
information alone is not a viable to competitor to other methods which 
use additional osbervables. For example, Huff et al.~(2013) have
shown that spectroscopic information can in principle
reduce the effect of shape
noise on both components of shear by an order of magnitude, rendering it
negible, whereas we have only shown reduction by a few tens of percent.
On the other hand, the photometric information will be present in 
catalogues without additional effort, so that using it 
should at least be considered.

In our work there are two
main distinctions between parameters, whether they are affected by
lensing (e.g. size), or are unaffected (e.g. colour).
A prediction of ellipticities from the latter
parameters has the advantage that the predicted ellipticity should 
not be affected by 
lensing. There should therefore be no correlation between the weak
lensing shear that is eventually measured after using the predicted
ellipticity, and the predicted ellipticity itself. This purity, as
we have seen, does come at  significant cost to the predictive power, and so
it becomes necessary to consider the more inclusive set of parameters,
which does not exclude those affected by lensing. In this case, because one can
regard our prediction of ellipticities as being to first order, one
might expect the effect of weak lensing on the parameters that 
enter into the prediction to modify the resulting predicted
ellipticities only at second order. We therefore expect that 
the effect of lensing on the prediction should be small. We defer
the developments of techniques to address this further to the future.

In this paper, we have also left to future work to explore how
best the predicted ellipticity information can be incorporated into
an estimator of the weak lensing shear. When this
is done, the fact that ellipticity predictions from photometry only extend to
galaxy shapes and not position angles, thus restricting any benefits
to only one component of the shear should also be taken into account.
It is possible that the predictions
are also better for certain subsets of the data (e.g.~bright galaxies)
and this could also be explored. 

One potential complication which could conceivably affect the reliability of
the techniques in this paper is
that there may be environmental effects on the relationship between
photometric parameters and predicted ellipticities. This would manifest
itself as spatial clustering in the residuals of the relationship, and could
cause systematic errors in the inferred shear. The magnitude of such effects
could perhaps be gauged using measures of the environment (e.g.~$n$th 
nearest neighbour distance). Spatial correlations in residuals from the
Fundamental Plane (FP) relationship between photometric and spectroscopic
parameters of early-type galaxies have recently been detected (Joachimi
et al.~2015), showing that such effects are present in related data.

\section*{Acknowledgments}
We thank Eric Huff, Melanie Simet and Rachel Mandelbaum for useful discussions.
This work is based on observations obtained with MegaPrime/MegaCam, a
joint project of CFHT and CEA/DAPNIA, at the Canada-France-Hawaii
Telescope (CFHT) which is operated by the National Research Council
(NRC) of Canada, the Institut National des Sciences de l'Univers of
the Centre National de la Recherche Scientifique (CNRS) of France, and
the University of Hawaii. This research used the facilities of the
Canadian Astronomy Data Centre operated by the National Research
Council of Canada with the support of the Canadian Space Agency.
CFHTLenS data processing was made possible thanks to significant
computing support from the NSERC Research Tools and Instruments grant
program.

{}

\end{document}